# INTERACTION OF NEUTRALINO DARK MATTER WITH COSMIC RAYS AND PAMELA/ATIC DATA

A.B. Flanchik

Institute of Radio Astronomy of NAS of Ukraine
e-mail: alex.svs.fl@gmail.com

The data obtained by PAMELA [1] and ATIC [2] show the presence of the peak in the cosmic positron spectrum at energies above 100 GeV. In this paper it has been shown that the peak can arise due to processes of the interaction of cosmic rays with neutralino dark matter, which are accompanied by chargino production and its leptonic decay.

1. Introduction. Usually it is believed that dark matter in the Universe must reveal itself due to gravity. However if it consists of light neutralinos $\tilde{\chi}_1^0$ then in addition to gravity such dark matter should participate in electroweak interactions. Earlier this problem was considered with respect to processes of neutralino annihilation into quarks, leptons, vector and Higgs bosons. This paper deals with problem of the cold neutralino dark matter interaction with electrons of cosmic rays $e^- + \tilde{\chi}_1^0 \to \nu_e + \tilde{\chi}_1^-$, with $\tilde{\chi}_1^-$ being a chargino. As it will be shown the leptonic decay of the final chargino $\tilde{\chi}_1^- \to \tilde{\chi}_1^0 + W^- \to \tilde{\chi}_1^0 + l^- + \bar{\nu}_l$ gives a signal in the form of the lepton-antineutrino pair with energies in the fixed interval. In the spectrum of cosmic leptons and antineutrino in this interval one should detect a peak similar to one which results from PAMELA and ATIC data.

2. The process $e^- + \tilde{\chi}_1^0 \to \nu_e + \tilde{\chi}_1^-$. Using energy and momentum conservation laws and considering a neutralino as resting we obtain for energies of final neutrino and chargino [3]

$$\omega_2 = \frac{m_0^2 - m^2 + 2m_0\omega_1}{2[m_0 + \omega_1(1-\cos\vartheta)]}, \quad \varepsilon_2 = \frac{m^2 + m_0^2 + 2\omega_1(m_0 + \omega_1)(1-\cos\vartheta)}{2[m_0 + \omega_1(1-\cos\vartheta)]}, \quad (1)$$

where $m_0, m$ are the neutralino and chargino masses, $\omega_1$ is the initial electron energy, $\vartheta$ is an angle between the electron and neutrino momenta. The process cross section as a function of the angle $\vartheta$ has a form [3]

$$\frac{d\sigma}{do} = \frac{G_F^2 m_W^4}{\pi^2} \left( \frac{m_0^2 - m^2 + 2m_0\omega_1}{m_W^2[m_0 + \omega_1(1-\cos\vartheta)] + \omega_1(m_0^2 - m^2 + 2m_0\omega_1)(1-\cos\vartheta)} \right)^2 \times$$

$$\left\{ |a|^2 + |b|^2 \frac{2m_0[m_0 + \omega_1(1-\cos\vartheta)] - (m_0^2 - m^2 + 2m_0\omega_1)(1-\cos\vartheta)}{2[m_0 + \omega_1(1-\cos\vartheta)]^2} - \right.$$

$$\left. \text{Re}(a^*b) \frac{m(1-\cos\vartheta)}{m_0 + \omega_1(1-\cos\vartheta)} \right\}, \quad do = 2\pi\sin\vartheta d\vartheta, \quad (2)$$

where $G_F$ is the Fermi coupling constant, $m_W$ is the W-boson mass, a and b are constants which determinate the chargino-neutralino charge current [4]. In Eq. (2) we take into account only the contribution of the W-boson exchange diagram, because it dominates at high energies. The angular distribution (2) has a maximum at small angles $\vartheta \ll 1$. In this limit from Eq. (1) we have

$$\omega_2 = \frac{m_0^2 - m^2 + 2m_0\omega_1}{2m_0}, \quad \varepsilon_2 = \frac{m^2 + m_0^2}{2m_0}. \tag{3}$$

Due to relativistic aberration the final neutrino takes away most of the electron energy and the chargino energy tends to fixed limit which does not depend on the electron energy. This situation is analogous to the 255.5 keV line formation at ultrarelativistic positron annihilation $e^- + e^+ \to 2\gamma$ with an electron at rest [5], when the one of the final photons carries almost all the positron energy and is radiated along the positron momentum direction while the other photon is emitted in opposite direction and has a fixed energy $m_e/2 = 255.5 \, keV$, where $m_e$ is the electron mass.

3. Two-body decay of a relativistic chargino. In case of the two-body decay $\tilde{\chi}_1^- \to \tilde{\chi}_1^0 + W^-$ the W-boson energy takes values in the range

$$E_{\min}(\varepsilon_1) \leq E \leq E_{\max}(\varepsilon_1),$$

$$E_{\min} = \varepsilon_1 \left(1 - \frac{\varepsilon_0 + v_1\sqrt{\varepsilon_0^2 - m_0^2}}{m}\right), \quad E_{\max} = \varepsilon_1 \left(1 - \frac{m_0^2}{m\varepsilon_0}\right), \tag{4}$$

where $\varepsilon_1$ is the chargino energy, $v_1 = \sqrt{1 - m^2/\varepsilon_1^2}$ is its velocity, $\varepsilon_0 = (m^2 + m_0^2 - m_W^2)/(2m)$. Here and below it is assumed that $v_1 > (m^2 - m_0^2 - m_W^2)/(m^2 + m_0^2 - m_W^2)$. The decay differential width reads

$$d\Gamma(\tilde{\chi}_1^- \to \tilde{\chi}_1^0 W^-) = \frac{m^2}{|\vec{p}_1|} \frac{\Gamma_\chi}{\sqrt{m^4 + (m_0^2 + m_W^2)^2 - 2m^2(m_0^2 + m_W^2)}} dE, \tag{5}$$

where $|\vec{p}_1| = \sqrt{\varepsilon_1^2 - m^2}$, and $\Gamma_\chi$ is the total chargino decay width [4]. Similarly the leptonic W-boson decay width as a function of the lepton energy $\omega$ is given by

$$d\Gamma(W^- \to l^- \bar{\nu}_l) = \frac{\Gamma_W}{q} d\omega, \quad q = \sqrt{E^2 - m_W^2}, \tag{6}$$

with $\Gamma_W$ being the total width of the decay $W^- \to l^- \bar{\nu}_l$. The energy $\omega$ for a fixed $E$ takes values from the range

$$\frac{m_W^2}{2(E+q)} \leq \omega \leq \frac{m_W^2}{2(E-q)}. \tag{7}$$

The distribution (7) has a form of a step and its height and width depend on the W-boson energy. With decreasing of the energy E the length of the value range of the energy $\omega$ decreases and the step height increases. Taking into account contributions of the all values of the energy $E$ from the range (4) one can obtain the lepton energy distribution which is given by the envelope in Fig. 1. Minimal and maximal energies of the lepton are determinate by Eqs. (7) and given by

$$\omega_{\min} = \frac{m_W^2}{2(E_{\max} + \sqrt{E_{\max}^2 - m_W^2})}, \quad \omega_{\max} = \frac{m_W^2}{2(E_{\max} - \sqrt{E_{\max}^2 - m_W^2})}. \tag{8}$$

For example we have $\omega_{\min} \approx 10$ GeV and $\omega_{\max} \approx 200$ GeV at the masses m = 250 GeV and m₀ = 110 GeV. Thus the decay processes $\tilde{\chi}_1^- \to \tilde{\chi}_1^0 + W^-$ finally leads to the peak in the cosmic lepton spectrum which results due to the leptonic decay $W^- \to l^- \bar{\nu}_l$. Let us notice that similar peaks in the spectrum of cosmic electrons, positrons and neutrinos can arise due to interaction processes $\nu_e + \tilde{\chi}_1^0 \to e^- + \tilde{\chi}_1^+$, $e^+ + \tilde{\chi}_1^0 \to \bar{\nu}_e + \tilde{\chi}_1^+$ with following chargino decay $\tilde{\chi}_1^+ \to \tilde{\chi}_1^0 + W^+ \to \tilde{\chi}_1^0 + l^+ + \nu_l$.

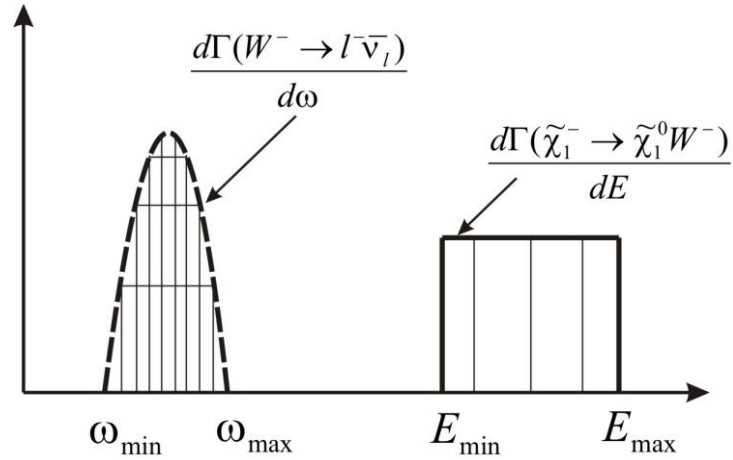

Fig. 1. Energy distributions of W-boson (right) and leptons arising from the W-boson decay (left).

4. Conclusions. In this paper the neutralino dark matter interaction with electrons of cosmic rays is considered in terms of the process $e^- + \tilde{\chi}_1^0 \to \nu_e + \tilde{\chi}_1^-$. It has been shown that at high energies of initial electrons the chargino energy tends to the fixed value $\varepsilon_{\tilde{\chi}} \approx (m^2 + m_0^2)/(2m_0)$, and the final neutrino takes away most of the electron energy. The chargino decay should help to detect the neutralino dark matter interactions with cosmic rays due to arising leptons and neutrinos with energies in the range from tens of GeV to several TeV depending on the neutralino and chargino masses. This signal from dark matter has to reveal itself as a peak in the energy spectrum of cosmic neutrinos and leptons and it looks like the peak detected by PAMELA and ATIC. An investigation of the spectrum of cosmic rays with energies up to several TeV is important for dark matter detection.

The author thanks V.M. Kontorovich, V.M. Shul'ga and D.P. Barsukov for useful discussions.

1. O. Adriani, G.C. Barbarino, G.A. Bazilevskaya et al., Nature, 458, 607 (2009), ArXiv: 0810.4995.
2. J. Chang, J.H. Adams, H.S. Ahn et al., Nature, 456, 362 (2008).
3. A.B. Flanchik, 25-th Conference High Energy Astrophysics Today and Tomorrow – 2010, Moscow, 21 – 24 Dec 2010, Abstract Book, p.54.
4. J.F. Gunion, H.E. Haber, R.M. Barnett et al., International Journal of Modern Physics A 2, 1145 (1987).
5. V.M. Kontorovich, O.M. Ulyanov, A.B. Flanchik, AIP Conference Proceedings 1269, 451 (2010).